\documentclass{osa-article}
\usepackage{optidef}
\usepackage{bm}
\usepackage{siunitx}
\usepackage{mathtools}
\usepackage{subcaption}
\usepackage{xcolor}
\usepackage{braket}
\usepackage{multirow}
\DeclareMathOperator{\Tr}{Tr}
\journal{osajournal}

\begin{document}

\title{Robust Polarimetry via Convex Optimization}

\author{Jacob M. Leamer,\authormark{1,2} Wenlei Zhang,\authormark{1,2} Ravi K. Saripalli, \authormark{1} Ryan T. Glasser, \authormark{1,*} and Denys I. Bondar\authormark{1, **}}

\address{\authormark{1}Department of Physics and Engineering Physics, Tulane University, 6823 St. Charles Avenue, New Orleans, LA 70118, USA\\
\authormark{2}These authors contributed equally to this work}

\email{\authormark{*}rglasser@tulane.edu}
\email{\authormark{**}dbondar@tulane.edu}

\begin{abstract}
We present mathematical methods, based on convex optimization, for correcting non-physical coherency matrices measured in polarimetry. We also develop the method for recovering the coherency matrices corresponding to the smallest and largest values of the degree of polarization given the experimental data and a specified tolerance. We use experimental non-physical results obtained with the standard polarimetry scheme and a commercial polarimeter to illustrate these methods. Our techniques are applied in post-processing, which compliments other experimental methods for robust polarimetry. 
\end{abstract}

\section{Introduction}

Polarization describes the trajectory of the electric field vector of light as it oscillates. Polarimetry and polarization imaging enable technologies in many fields, such as machine vision \cite{wolff_polarization-based_1990,koshikawa_model-based_1985}, remote sensing \cite{tyo_review_2006,schott_fundamentals_2009}, biomedical optics \cite{ghosh_tissue_2011}, astronomy \cite{vlemmings_2007,deglinnocenti_polarization_2004}, and free-space optical communication \cite{Zhang:17,Ma:17,10.1117/12.2312038}. Many quantum information protocols also depend on the determination of polarization states \cite{gasparoni_realization_2004,kok_linear_2007,obrien_photonic_2009,kagalwala_single-photon_2017}. A new generation of polarization imaging cameras is currently under development, which will further accelerate the application of polarimetry in many fields \cite{rubin_matrix_2019}. \par
The state of polarization can be described by the Stokes parameters \cite{stokes_2009,born_principles_1999} or the coherency matrix \cite{goodman_statistical_2000}, which is a generalization of the Jones calculus \cite{jones_new_1941}. The Stokes parameters, $s_0,s_1,s_2,s_3$, and the coherency matrix, $\mathbf{J}$, are related by

\begin{gather}
    \mathbf{J}=\frac{1}{2}\begin{pmatrix}
    s_0+s_1 & s_2+is_3 \\
    s_2-is_3 & s_0-s_1
    \end{pmatrix}. \label{eq:jones}
\end{gather}
The coherency matrix provides all second-order statistical information about the polarization state. \par
As shown in later sections, non-physical coherency matrices (for example, with negative eigenvalues) arise in common polarimetry schemes due to experimental errors, such as fluctuations of the light source, imperfect alignment of optical components, and spectral bandwidth of the source.  Several techniques based on pre-processing, calibration, estimation of the Stokes parameters \cite{Zallat:s,valenzuela_joint_recon,Paxman:s,faisan_filtering_est} and in-situ optimization\cite{foreman_priori_2008,zhi_error_2017, shemirani_Compensation_2010} or novel polarimetric schemes \cite{shi_automatic_2006,collins_galway_2013,lopez-mago_overall_2019,clarke_interference_2004} have been developed to reduce the effect of such experimental errors. One such method of particular interest is the constrained maximum likelihood (CML) method \cite{hu_cml_2013} due to its similarities to our own work in terms of both motivation and application.\par
In this article, we present a method of correcting these non-physical results by finding the closest physical coherency matrix via convex optimization. This method is applied in post-processing, and does not depend on \textit{a priori} information or the experimental setup. Having such a method is especially useful when dealing with other degrees of freedom in addition to polarization. For example, when measuring the polarization profile of a vector beam, we might have the result where only a few points are non-physical due to experimental errors \cite{suzuki_comprehensive_2019}. Using our method, we can find the closest approximate physical coherency matrix for these points rather than invalidating all points and repeating the entire measurement. It is also potentially useful when dealing with measurements that cannot be easily repeated, such as the polarimetry of single photons. This method can be easily generalized to be used for multi-photon Stokes parameters \cite{abouraddy_quantum_2002}.

\section{Experimental Setup} \label{sec:setup}

\begin{figure}
    \centering
    \includegraphics[width=0.5\linewidth]{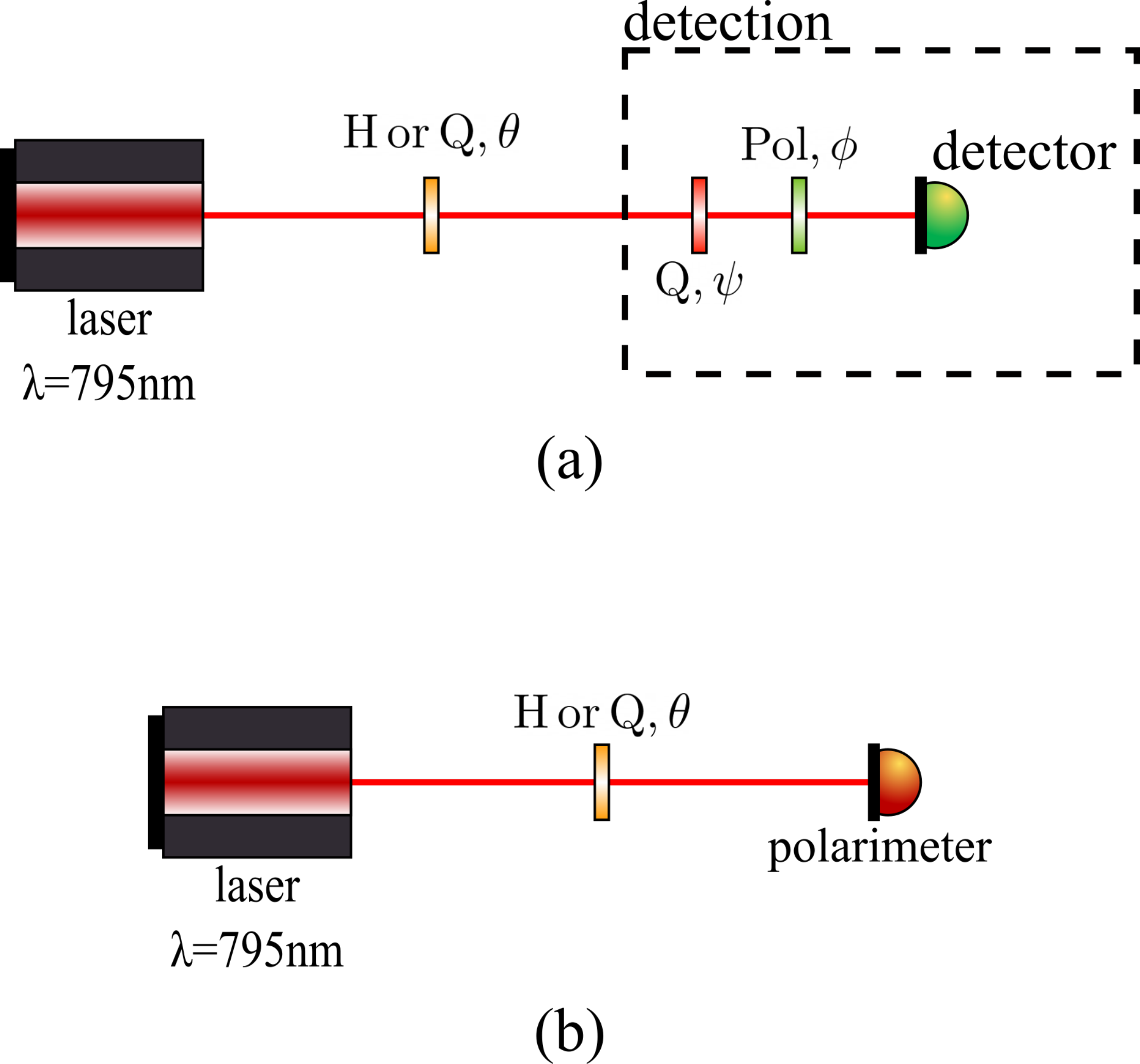}
    \caption{Experimental polarimetry setup: (a) modified standard method; (b) polarimeter method. H: half-wave plate; Q: quarter-wave plate; Pol: linear polarizer.}
    \label{fig:setup_polarimetry}
\end{figure}

We used the two independent polarimetry schemes shown in Fig.~\ref{fig:setup_polarimetry} to measure the coherency matrices of both linearly and elliptically polarized light to verify the validity of the developed methods. In both schemes, the light from the laser is vertically polarized, and passes through either a half-wave plate (HWP) or a quarter-wave plate (QWP). The HWP preserves the linearity of the laser light, but changes the angle of polarization, while the QWP changes linear polarization to elliptical polarization. The exact polarization state after the waveplates depends on $\theta$, which is the angle between the fast axis of the waveplates and the horizontal axis. In Fig.~\ref{fig:setup_polarimetry} (a), we use a modified version of the standard method for measuring the Stokes parameters \cite{born_principles_1999,collett_polarized_1993}. The detection scheme consists of another QWP, a linear polarizer, and an intensity detector. The following four intensity measurements are required to measure all four Stokes parameters: $I(0^\circ,0^\circ)$, $I(0^\circ,90^\circ)$, $I(0^\circ,45^\circ)$, and $I(45^\circ,45^\circ)$, where $I(\psi,\phi)$ is the intensity measured by the detector when the fast axis of the QWP (in the detection scheme) is at angle $\psi$ w.r.t. the horizontal axis and the axis of transmission of the polarization is at angle $\phi$ w.r.t. the horizontal axis. The Stokes parameters can be calculated from the intensity measurements using the following equations,
\begin{gather}
    s_0=I(0^\circ,0^\circ)+I(0^\circ,90^\circ), \label{EqS0}\\
    s_1=I(0^\circ,0^\circ)-I(0^\circ,90^\circ), \\
    s_2=2I(45^\circ,45^\circ)-s_0, \\
    s_3=2I(0^\circ,45^\circ)-s_0. \label{EqS3}
\end{gather} \par
Care was taken to ensure proper alignment of the quarter-wave-plate and linear polarizer. However, only four data points are measured, which increases the chance non-physical results due to fluctuations of the laser light between the four intensity measurements. In the polarimeter method (Fig.~\ref{fig:setup_polarimetry} (b)), the measurement is done solely with a polarimeter, and the Stokes parameters are given automatically. The polarimeter employs a spinning waveplate and curve-fitting technique to obtain the Stokes parameters \cite{schaefer_measuring_2007}. While this method is fast (sampling rate up to \SI{400}{Hz}), the polarimeter still sometimes produces non-physical results especially if the input power exceeds 2 mW as per company specification. In both methods different input states were prepared using a quarter wave plate and rotated by 10 degrees for one half of a complete rotation which is sufficient to obtain a complete cycle of intensity. Using both methods, we measured the Stokes parameters of high intensity light ($\sim\SI{16}{mW}$) and low intensity light  ($\sim\SI{1.4}{mW}$) for both linearly and elliptically polarized light. Measurements with high input power of about 15 mW was used to produce non-physical results using the polarimeter.

\section{Polarimetry as a Convex Optimization Problem} \label{sec:method}

As stated previously, experimental errors can lead to non-physical results where the coherency matrix has negative eigenvalues. Such non-physical coherency matrices result in degrees of polarization (DOP) greater than 1. The DOP quantifies the portion of the light that is polarized, and it can be obtained from the Stoke parameters or the coherency matrix using \cite{born_principles_1999}
\begin{gather}\label{eq_DOP_def}
    \mathrm{DOP}^2=\frac{s_1^2+s_2^2+s_3^2}{s_0^2}= \frac{2}{s_0^2}\Tr(\mathbf{J}^2)-1.
\end{gather}

Throughout the paper, the measured coherency matrix, $\mathbf{J}_\mathrm{measured}$, is constructed from the Stokes parameters in Eqs.~\eqref{EqS0}--\eqref{EqS3} using Eq.~\eqref{eq:jones} and normalized such that $\operatorname{Tr}(\mathbf{J}_\mathrm{measured})=1$.

We want to find the corrected coherency matrix, $\mathbf{J}_{\mathrm{corrected}}$, that is closest to $\mathbf{J}_{\mathrm{measured}}$ under the constraint that $\mathbf{J}_{\mathrm{corrected}}$ be physical.  This is equivalent to solving the following optimization problem,
\begin{mini!}|l|
    {\mathbf{J}_{\mathrm{corrected}}}{\|\mathbf{J}_{\mathrm{measured}} - \mathbf{J}_{\mathrm{corrected}}\|}{}{} \label{objective}
    \addConstraint{\mathbf{J}_{\mathrm{corrected}}}{\geq 0}{} \label{constraint1}
    \addConstraint{\Tr(\mathbf{J}_{\mathrm{corrected}})}{= 1}{}. \label{constraint2}
\end{mini!}

This problem is an example of a convex optimization problem, which is an optimization problem where the objective and constraint functions $f_i(x): \mathcal{C} \rightarrow \mathbb{R}$ are convex, i.e. they satisfy the condition that for all $x,y \in \mathcal{C}$, 
\begin{equation} 
    f_{i}(t x + (1-t)y) \leq t f_{i}(x) + (1-t) f_i(y),
    \label{convex}
\end{equation} 
where $t \in[0,1]$ and $\mathcal{C}$ is some convex set \cite{boyd_vandenberghe_2004}.  A convex set $\mathcal{C}$ is defined as a set where, for all $x,y \in \mathcal{C}$ and $t \in [0,1]$, $t x + (1-t)y \in \mathcal{C}$.  The constraint presented in Eq.~\eqref{constraint1} restricts the possible choices for $\mathbf{J}_{\mathrm{corrected}}$ to the set of positive semi-definite matrices, which are known to form a convex set \cite{neumann}.  The objective function Eq.~\eqref{objective} in our problem is convex due to the definition of norms, namely that they are subadditive and absolutely scalable.  Finally, we know that Eq.~\eqref{constraint2} is convex because of the linearity of trace, $\Tr(\alpha \mathbf{A} + \beta \mathbf{B}) = \alpha \Tr(\mathbf{A}) + \beta \Tr(\mathbf{B})$, which satisfies Eq.~\eqref{convex} with equality.  Thus, our problem is convex.  Since the solution to a convex optimization problem is unique and provides a lower bound on more general optimization problems, the ability to construct and solve a convex optimization problem has proven useful in a wide variety of topics such as the reconstruction of quantum channels \cite{huang_2020}, the selection of sensors to minimize error in a measurement \cite{joshi_boyd_2009}, and multi-period trading \cite{boyd_busseti_diamond_kahn_koh_nystrup_speth_2017}.  As such, a number of tools and techniques for solving convex optimization problems efficiently have been developed. In particular we opted to use Matlab's CVX library\cite{cvx,gb08} due to its ability to handle complex matrices and its ease of use compared to other options.  \par

\begin{figure}[htbp]
    \centering
    \includegraphics[width=.8\linewidth]{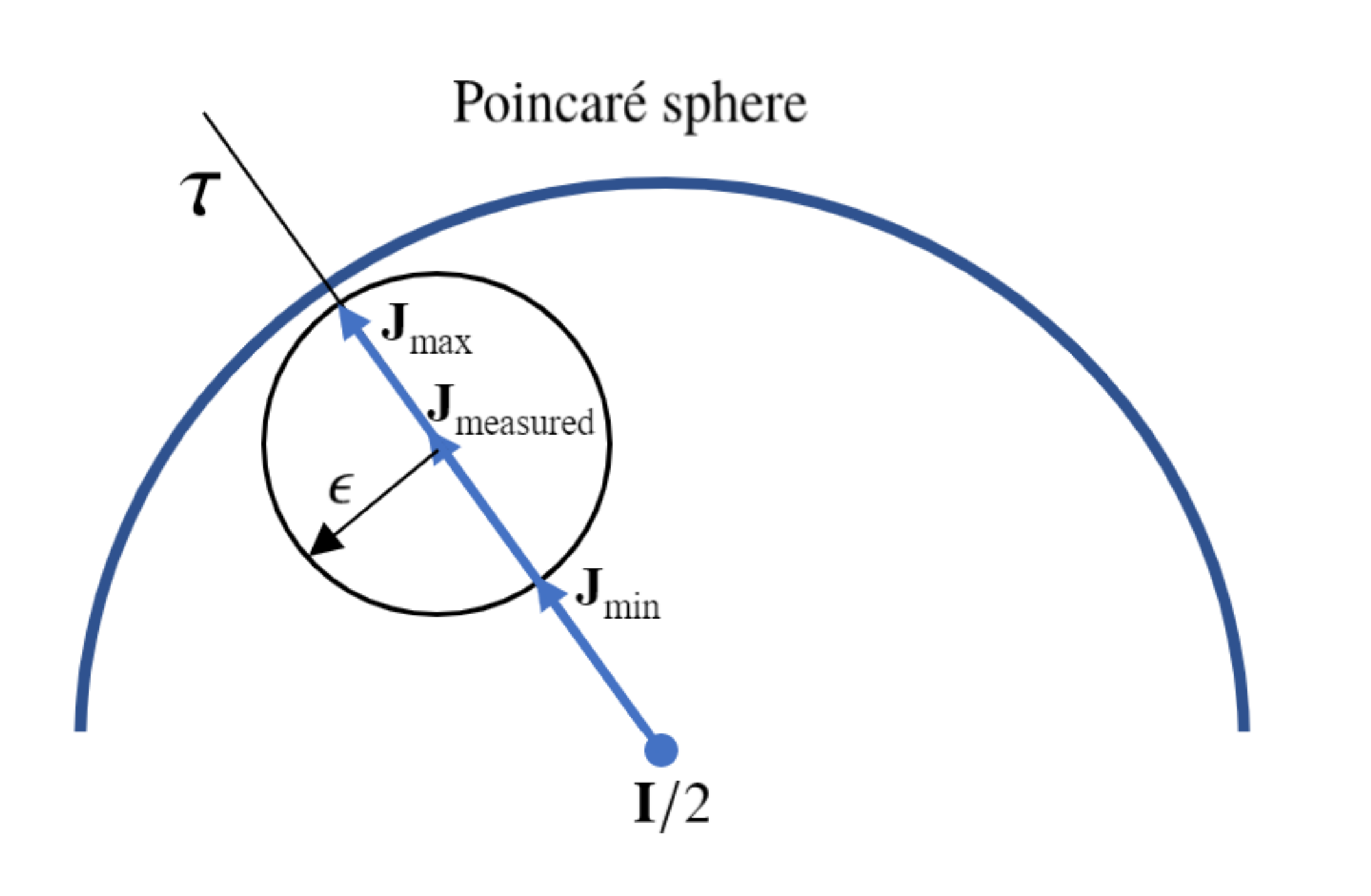}
    \caption{Schematic for the construction of a convex optimization problem to determine the coherency matrices with the highest and lowest DOP (respectively $\mathbf{J}_{\mathrm{max}}$ and $\mathbf{J}_{\mathrm{min}}$) for a given $\mathbf{J}_{\mathrm{measured}}$ and error tolerance $\epsilon$. The blue arrows are the Stokes vectors of the corresponding coherency matrices. $\mathbf{I}/2$ corresponds to the zero vector, where $\mathbf{I}$ is the identity matrix.}
    \label{fig:schematic}
\end{figure}

We also want to determine the upper and lower bounds on the DOP for $\mathbf{J}_{\mathrm{corrected}}$ given some tolerance $\epsilon$ for the acceptable difference between $\mathbf{J}_{\mathrm{corrected}}$ and $\mathbf{J}_{\mathrm{measured}}$. The appropriate choice for the value of $\epsilon$ will be determined by the amount of noise in the measurement of $\mathbf{J}_{\mathrm{measured}}$.  If there is a lot of noise, a larger value must be selected for the tolerance in order to obtain solutions that still lie within the Poincar\'e sphere.  Let $\mathbf{J}_{\min}$ denote the value for $\mathbf{J}_{\mathrm{corrected}}$ with the lowest DOP for the specified tolerance $\epsilon$; likewise, $\mathbf{J}_{\max}$ be the value for $\mathbf{J}_{\mathrm{corrected}}$ with the largest DOP. If we consider the representation of coherence matrices on the Poincar\'{e} sphere as shown in Fig.~\ref{fig:schematic}, the Stokes vectors for  $\mathbf{J}_{\min}$ and  $\mathbf{J}_{\max}$ must lie within a ball of radius $\epsilon$ centered on $\mathbf{J}_{\mathrm{measured}}$. With this information and Eq.~\eqref{eq_DOP_def}, we might be tempted to construct convex optimization problems as 
\begin{mini}|l|
    {\mathbf{J}_{\min}}{\Tr(\mathbf{J}_{\min}^2)}{}{}
    \addConstraint{\|\mathbf{J}_{\min}-\mathbf{J}_{\mathrm{measured}}\|}{\leq \epsilon}{}
    \addConstraint{\Tr(\mathbf{J}_{\min})}{= 1}{}
    \addConstraint{\mathbf{J}_{\min}}{\geq 0}{}
    \label{naive_min}
\end{mini}
for finding $\mathbf{J}_{\min}$ or
\begin{maxi}|l|
    {\mathbf{J}_{\max}}{\Tr(\mathbf{J}_{\max}^2)}{}{}
    \addConstraint{\|\mathbf{J}_{\max}-\mathbf{J}_{\mathrm{measured}}\|}{\leq \epsilon}{}
    \addConstraint{\Tr(\mathbf{J}_{\max})}{= 1}{}
    \addConstraint{\mathbf{J}_{\max}}{\geq 0}{}
    \label{naive_max}
\end{maxi}
for finding $\mathbf{J}_{\max}$, but attempting to solve either of these problems is actually impossible using MATLAB's CVX library.  This is due to the rules (called the Disciplined Convex Programming (DCP) ruleset) that the library uses to determine if an expression is a valid convex optimization problem.  In particular, the library cannot determine whether the objective functions of Eq.~\eqref{naive_min} or Eq.~\eqref{naive_max} are convex because they involve the squaring of variables that are not scalars.  However, if we consider other ways of characterizing the DOP, it is possible to reformulate Eq.~\eqref{naive_min} or Eq.~\eqref{naive_max} so that they satisfy the DCP rules.  By Eq.~\eqref{eq_DOP_def}, we also know that the DOP of a coherency matrix corresponds to the length of its Stokes vector.  If a ray (labelled $\tau$ in Fig.~\ref{fig:schematic}) is drawn from the origin of the Poincar\'{e} sphere in the direction of the Stokes vector for $\mathbf{J_\mathrm{measured}}$, the parameter $\tau$ can be introduced to describe the length of Stokes vectors that lie on the ray such that a larger value for $\tau$ describes a longer Stokes vector.  This means that $\tau$ also describes the DOP of the coherency matrices that correspond to these Stokes vectors, and that $\mathbf{J}_{\mathrm{max}}$ and $\mathbf{J}_{\mathrm{min}}$ are found where this ray intersects the $\epsilon$-ball.  Thus, we require that the sought $\mathbf{J}_{\mathrm{max}}$ and $\mathbf{J}_{\mathrm{min}}$ lie on the ray using the expression,
\begin{equation}
    \left\|\mathbf{J}_{\min, \max} - \tau\mathbf{J}_{\mathrm{measured}} - \frac{1-\tau}{2}\mathbf{I}\right\| = 0,
    \label{convexline}
\end{equation}
so that $\mathbf{J}_{\max}$ and $\mathbf{J}_{\min}$ can be found by maximizing and minimizing $\tau$, respectively. Our objective function is now just $\tau$, which is clearly a convex function, and it will satisfy the DCP rules of MATLAB's CVX library. It is important to note that the value of $\tau$ obtained via optimization not only depends upon $\mathbf{J}_{\mathrm{measured}}$ and $\epsilon$, but also on the constraint both $\mathbf{J}_{\min, \max}$ be non-negative as in the problem~\eqref{objective}--~\eqref{constraint2}. 
Because the l.h.s. of Eq.~\eqref{convexline} is a norm of a linear expression over a convex set, the constraint function \eqref{convexline} is convex.  Now we can combine the constraints~\eqref{constraint1}, ~\eqref{constraint2}, and ~\eqref{convexline} with the requirement that $\mathbf{J}_{\min, \max}$ be $\epsilon$-close to $\mathbf{J}_{\mathrm{measured}}$ to formulate the following convex optimization problems for obtaining physically corrected coherency matrices $\mathbf{J}_{\min, \max}$ with the minimal and maximal DOP:

To recover the physically-constrained coherency matrix $\mathbf{J}_{\min}$ with the minimal DOP that is $\epsilon$-close to the measured  $\mathbf{J}_{\mathrm{measured}}$, we solve the convex optimization problem
\begin{mini}|l|
    {\tau, \mathbf{J}_{\min}}{\tau}{}{}
    \addConstraint{\|\mathbf{J}_{\min} - \tau\mathbf{J}_{\mathrm{measured}} - \frac{1-\tau}{2}\mathbf{I}\|}{= 0}{}
    \addConstraint{\|\mathbf{J}_{\min}-\mathbf{J}_{\mathrm{measured}}\|}{\leq \epsilon}{}
    \addConstraint{\Tr(\mathbf{J}_{\min})}{= 1}{}
    \addConstraint{\mathbf{J}_{\min}}{\geq 0}{}.
    \label{min_dop}
\end{mini}
Likewise, to obtain the physically-constrained coherency matrix $\mathbf{J}_{\max}$ with the maximal DOP that is $\epsilon$-close to the measured  $\mathbf{J}_{\mathrm{measured}}$, we solve
\begin{maxi}|l|
    {\tau, \mathbf{J}_{\max}}{\tau}{}{}
    \addConstraint{\|\mathbf{J}_{\max} - \tau\mathbf{J}_{\mathrm{measured}} - \frac{1-\tau}{2}\mathbf{I}\|}{= 0}{}
    \addConstraint{\|\mathbf{J}_{\max}-\mathbf{J}_{\mathrm{measured}}\|}{\leq \epsilon}{}
    \addConstraint{\Tr(\mathbf{J}_{\max})}{= 1}{}
    \addConstraint{\mathbf{J}_{\max}}{\geq 0}{}.
    \label{max_dop}
\end{maxi}

\section{Results} \label{sec:results}

\begin{figure}
    \centering
    \includegraphics[width=1\linewidth]{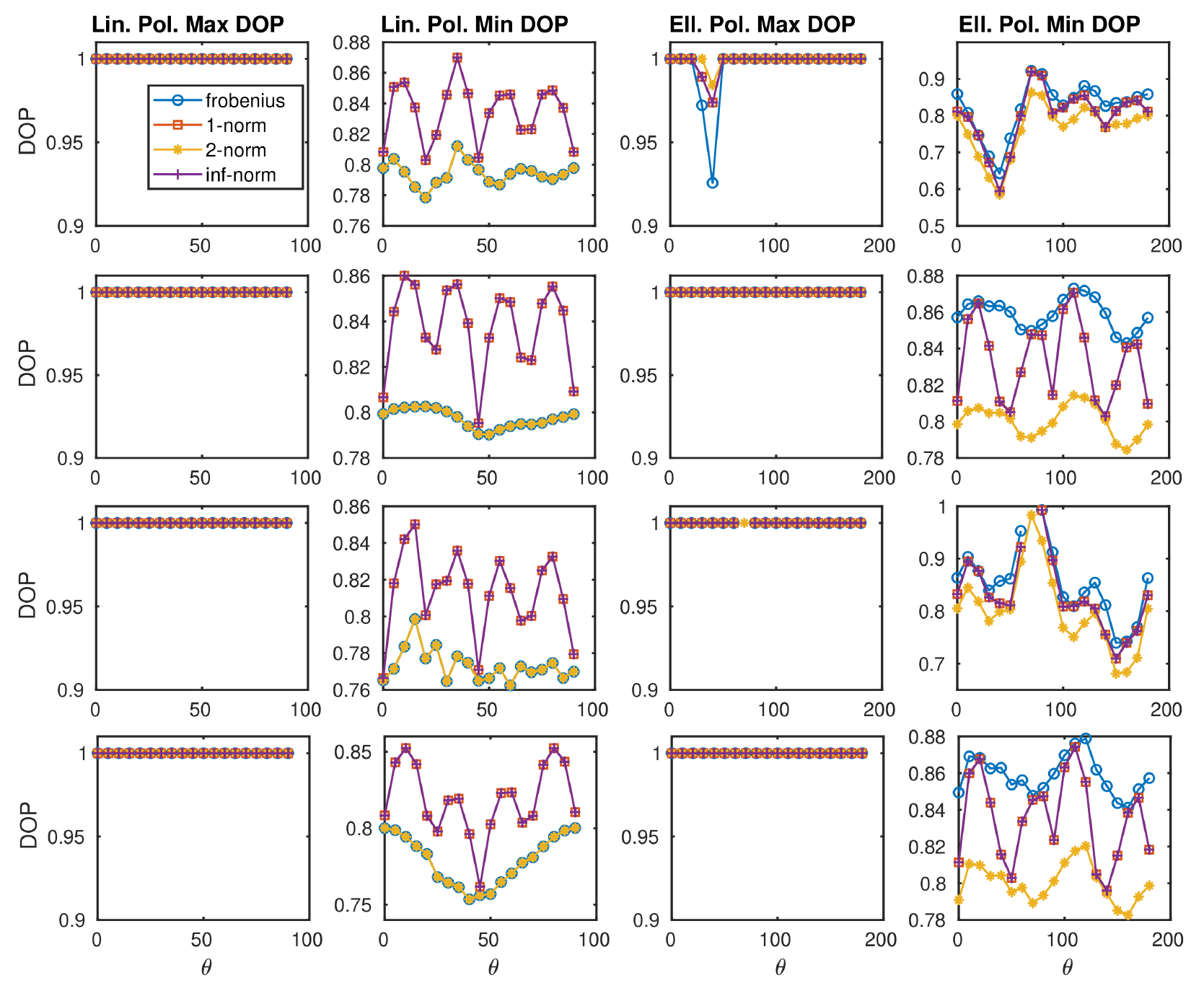}
    \caption{DOP of $\mathbf{J}_{\mathrm{\max}}$ and $\mathbf{J}_{\mathrm{\min}}$ for linearly polarized light (first and second columns respectively) and for ellptically polarized light (third and fourth columns respectively) obtained by solving Eqs.~\eqref{min_dop} and ~\eqref{max_dop} using the Frobenius norm, the 1-norm, the 2-norm, and the $\infty$-norm.  The first and second rows used high intensity light measured with the standard method and the polarimeter method respectively.  The third and fourth rows used low intensity light measured using the standard method and polarimeter method respectively.}
    \label{fig:norm}
\end{figure}

To solve Eqs.~\eqref{objective}--~\eqref{constraint2}, ~\eqref{min_dop}, and ~\eqref{max_dop}, we had to specify a norm.  We chose to use the Frobenius norm, which is defined by
\begin{equation}
    \left\|\mathbf{A}\right\|_\mathrm{F} = \sqrt{\Tr(\mathbf{A}^\dag\mathbf{A})},
\end{equation}
because, according to Eq.~\eqref{eq_DOP_def}, the Frobenius norm of a coherency matrix is related to its DOP.  While the use of the Frobenius norm may have been the natural choice given our interest in the DOP, the convex optimization problems outlined in Eqs.~\eqref{objective}--~\eqref{constraint2}, ~\eqref{min_dop}, and ~\eqref{max_dop} can be solved using any norm. Certain experimental applications may find that the use of a particular norm is more beneficial than others, however.  As an example, the use of the 1-norm may lead to better results for compressed-sensing applications.  Additionally, different norms may be useful for modelling different kinds of noise present in the system.  A comparison of the results of the minimization and maximization methods using the Frobenius norm, the 1-, $\infty$-, and 2-norms are displayed in Fig.~\ref{fig:norm}.  For an $m$-by$n$ matrix $\mathbf{A}$, these norms are defined in the following way
\begin{gather}
    \|\mathbf{A}\|_{1} = \max_{1 \leq j \leq n} \sum_{i=1}^m |a_{ij}|, \qquad
    \|\mathbf{A}\|_{\infty} = \max_{1 \leq i \leq m} \sum_{j=1}^n |a_{ij}|, \qquad
    \|\mathbf{A}\|_{2} = \sigma_{\max}(\mathbf{A}) \leq \|\mathbf{A}\|_\mathrm{F},
\end{gather}
where $\sigma_{\max}(\mathbf{A})$ is the largest singular value of $\mathbf{A}$.  As can be seen in Fig.~\ref{fig:norm}, the 1- and $\infty$-norms always give the same result for the DOP.  This is because the coherency matrix is a $2\times2$ symmetric matrix, which means that the maximum absolute column sum will always equal the maximum absolute row sum.  For linearly polarized light, the values of DOP obtained from the minimization problem Eq.~\eqref{min_dop} using the 1-norm (or $\infty$-norm) are always greater than those obtained using the Frobenius norm or the 2-norm.  Additionally, using either the 2-norm or the Frobenius norm gives the same results for linearly polarized light.  For elliptically polarized light, the results obtained using the Frobenius norm are higher than those obtained using the 2-norm, with the plots of the two forming upper and lower bounds on the results obtained using the 1-norm ($\infty$-norm).\par
\begin{figure}
    \centering
    \includegraphics[width=.95\linewidth]{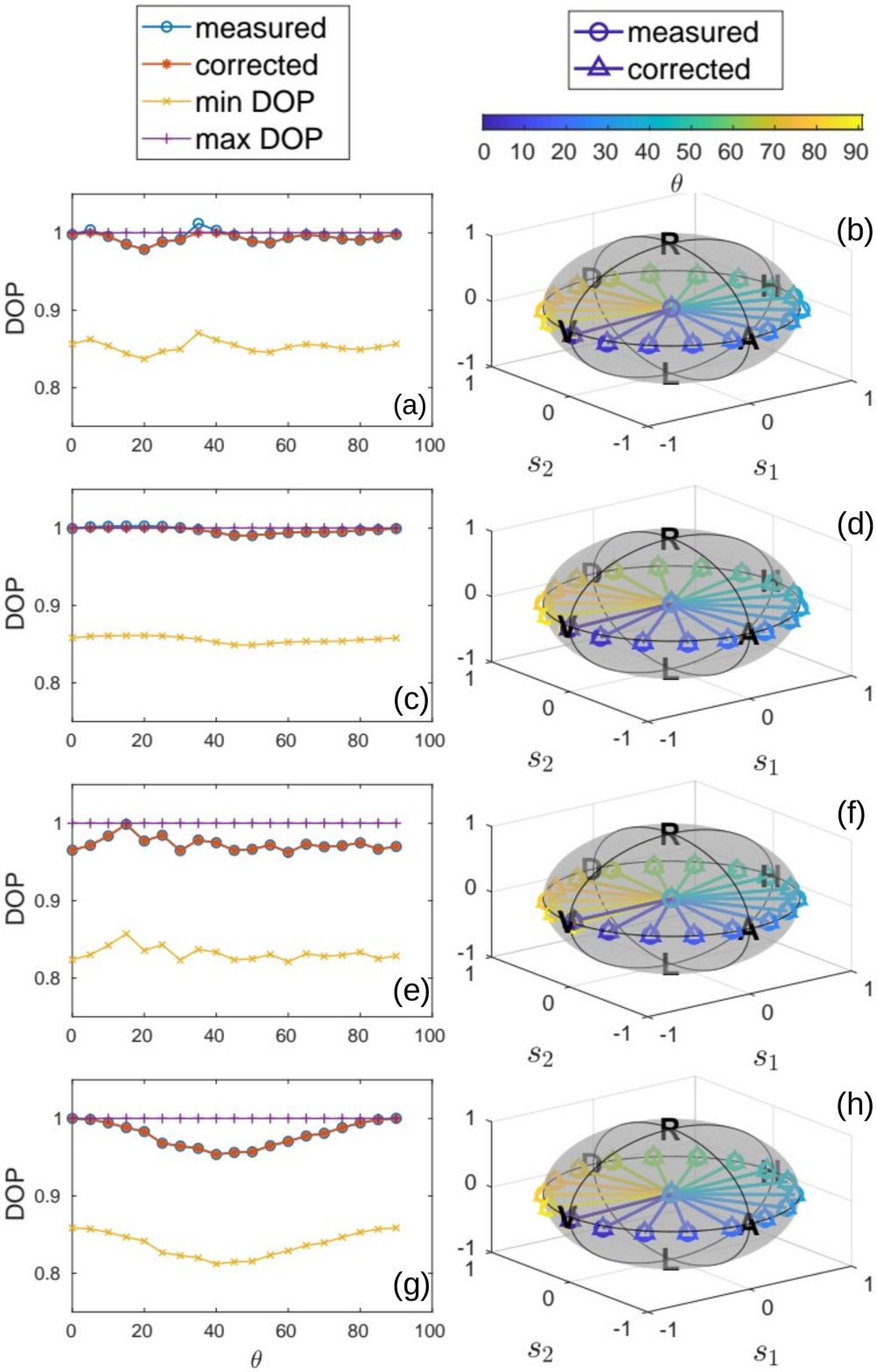}
    \caption{DOP of $\mathbf{J}_{\mathrm{measured}}$ and $\mathbf{J}_{\mathrm{corrected}}$ (left column) and the location of their corresponding vectors on the Poincar\'{e} sphere (right column) for linearly polarized light: (a), (b) high intensity light measured with the standard method; (c), (d) high intensity light measured with the polarimeter method; (e), (f) low intensity light measured with the standard method; (g), (h) low intensity light measured with the polarimeter method.  In each case, $\mathbf{J}_{\mathrm{corrected}}$ was obtained by solving Eqs.~\eqref{objective}--~\eqref{constraint2}.  The minimum and maximum DOP were calculated from the solutions to Eqs.~\eqref{min_dop} and ~\eqref{max_dop} using a tolerance of $\epsilon = 0.1$.}
    \label{fig:lp_fig}
\end{figure}
\begin{table}[htbp]
    \centering
    \begin{tabular}{|c|c|c|c|c|c|c|c|}
        \multicolumn{8}{c}{Linearly Polarized Data} \\
        \hline
        & \multicolumn{4}{|c|}{Measured}  & \multicolumn{3}{|c|}{Corrected} \\
        \hline
        Data Set& $\theta$ & $s_1/s_0$ & $s_2/s_0$ & $s_3/s_0$ & $s_1/s_0$ & $s_2/s_0$ & $s_3/s_0$ \\
        \hline
        \multirow{3}{4em}{High Intensity Standard} & 5 & -0.928 & -0.382 & -0.033 & -0.924 & -0.381 & -0.033 \\
        & 35 & 0.781 & -0.644 & -0.002 & 0.772 & -0.636 & -0.002 \\
        & 40 & 0.945 & -0.335 & -0.005 & 0.942 & -0.334 & -0.005 \\
        \hline
        \multirow{6}{4em}{High Intensity Polarimeter} & 5 & -0.947 & -0.325 & -0.038 & -0.945 & -0.325 & -0.038 \\
        & 10 & -0.771 & -0.639 & -0.038 & -0.770 & -0.637 & -0.038 \\
        & 15 & -0.503 & -0.866 & -0.039 & -0.502 & -0.864 & -0.038 \\
        & 20 & -0.199 & -0.982 & -0.039 & -0.199 & -0.979 & -0.039 \\
        & 25 & 0.160 & -0.989 & -0.036 & 0.160 & -0.987 & -0.036 \\
        & 30 & 0.494 & -0.869 & -0.033 & 0.494 & -0.869 & -0.033 \\
        \hline
        \multirow{2}{4em}{Low Intensity Polarimeter} & 0 & -0.999 & -0.010 & -0.043 & -0.999 & -0.010 & -0.043  \\ 
        & 90 & -0.999 & -0.005 & -0.056 & -0.998 & -0.005 & -0.056 \\
        \hline
    \end{tabular}
    \caption{Values of the Stokes parameters obtained from the experimental data (under the "Measured" column) and from the results of applying our convex optimization method (Eqs.~\eqref{objective}-~\eqref{constraint2}) (under the "Corrected" column) for linearly polarized light where the DOP $\geq$ 1 when calculated using the experimental data.}
    \label{tab:lp_data}
\end{table}
\begin{figure}
    \centering
    \includegraphics[width=.95\linewidth]{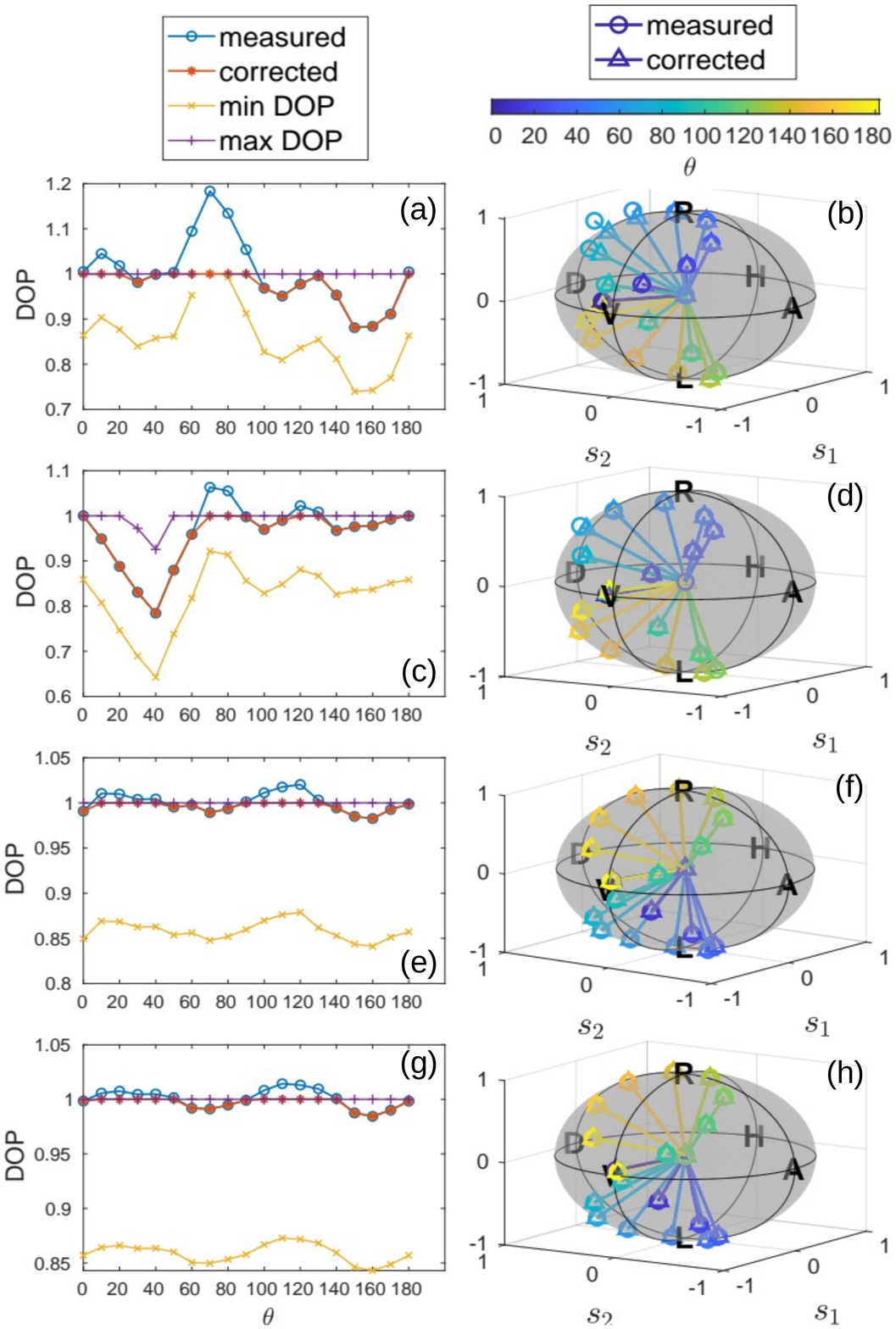}
    \caption{DOP of $\mathbf{J}_{\mathrm{measured}}$ and $\mathbf{J}_{\mathrm{corrected}}$ (left column) and the location of their corresponding vectors on the Poincar\'{e} sphere (right column) for elliptically polarized light: (a), (b) low intensity light measured with the standard method; (c), (d) high intensity light measured with the standard method; (e), (f) low intensity light measured with the polarimeter method; (g), (h) high intensity light measured with the polarimeter method.  In each case, $\mathbf{J}_{\mathrm{corrected}}$ was obtained by solving Eqs.~\eqref{objective}--~\eqref{constraint2}.  The minimum and maximum DOP were calculated from the solutions to Eqs.~\eqref{min_dop} and ~\eqref{max_dop} using a tolerance of $\epsilon = 0.1$.}
    \label{fig:cp_fig}
\end{figure}
\begin{table}[htbp]
    \small
    \centering
    \begin{tabular}{|c|c|c|c|c|c|c|c|}
        \hline
        \multicolumn{8}{|c|}{Elliptically Polarized Data} \\
        \hline
        \multicolumn{1}{|c|}{Data Set} & \multicolumn{4}{|c|}{Measured}  & \multicolumn{3}{|c|}{Corrected} \\
        \hline
        & $\theta$ & $s_1/s_0$ & $s_2/s_0$ & $s_3/s_0$ & $s_1/s_0$ & $s_2/s_0$ & $s_3/s_0$ \\
        \hline
        \multirow{9}{4em}{Low Intensity Standard} & 0 & -0.990 & 0.105 & 0.139 & -0.985 & 0.104 & 0.138 \\ 
        & 10 & -0.944 & -0.232 & 0.384 & -0.903 & -0.222 & 0.369 \\ 
        & 20 & -0.683 & -0.473 & 0.590 & -0.670 & -0.464 & 0.579 \\ 
        & 50 & -0.042 & 0.070 & 1.00 & -0.042 & 0.070 & 0.997 \\ 
        & 60 & -0.196 & 0.354 & 1.02 & -0.179 & 0.324 & 0.929 \\ 
        & 70 & -0.513 & 0.496 & 0.944 & -0.434 & 0.419 & 0.798 \\ 
        & 80 & -0.826 & 0.333 & 0.703 & -0.728 & 0.294 & 0.620 \\ 
        & 90 & -0.993 & 0.075 & 0.344 & -0.943 & 0.071 & 0.326 \\ 
        & 180 & -0.992 & 0.117 & 0.112 & -0.987 & 0.117 & 0.111 \\
        \hline
        \multirow{5}{4em}{High Intensity Standard} & 0 & -0.999 & 0.058 & 0.013 & -0.998 & 0.058 & 0.013 \\ 
        & 70 & -0.616 & 0.558 & 0.663 & -0.580 & 0.525 & 0.623 \\ 
        & 80 & -0.908 & 0.356 & 0.402 & -0.861 & 0.338 & 0.381 \\ 
        & 120 & -0.225 & -0.438 & -0.896 & -0.220 & -0.429 & -0.876 \\ 
        & 130 & -0.020 & -0.185 & -0.991 & -0.020 & -0.184 & -0.983 \\
        \hline
        \multirow{9}{4em}{Low Intensity Polarimeter} & 10 & -0.917 & -0.356 & -0.232 & -0.907 & -0.353 & -0.230 \\ 
        & 20 & -0.640 & -0.549 & -0.555 & -0.634 & -0.544 & -0.550 \\ 
        & 30 & -0.295 & -0.510 & -0.813 & -0.293 & -0.508 & -0.810 \\ 
        & 40 & -0.062 & -0.264 & -0.967 & -0.062 & -0.263 & -0.963 \\ 
        & 90 & -0.992 & -0.073 & -0.114 & -0.991 & -0.073 & -0.114 \\ 
        & 100 & -0.894 & -0.408 & 0.237 & -0.884 & -0.404 & 0.234 \\ 
        & 110 & -0.595 & -0.602 & 0.565 & -0.585 & -0.591 & 0.556 \\ 
        & 120 & -0.246 & -0.554 & 0.821 & -0.241 & -0.543 & 0.805 \\ 
        & 130 & -0.008 & -0.291 & 0.960 & -0.008 & -0.290 & 0.957 \\
        \hline
        \multirow{10}{4em}{High Intensity Polarimeter} & 10 & -0.904 & -0.342 & -0.278 & -0.899 & -0.340 & -0.277 \\ 
        & 20 & -0.610 & -0.529 & -0.602 & -0.606 & -0.525 & -0.598 \\ 
        & 30 & -0.262 & -0.474 & -0.847 & -0.261 & -0.471 & -0.843 \\ 
        & 40 & -0.032 & -0.223 & -0.979 & -0.031 & -0.222 & -0.975 \\ 
        & 50 & -0.019 & 0.110 & -0.995 & -0.019 & 0.110 & -0.994 \\ 
        & 100 &-0.877 & -0.399 & 0.299 & -0.869 & -0.396 & 0.296 \\ 
        & 110 & -0.578 & -0.569 & 0.609 & -0.570 & -0.561 & 0.600 \\ 
        & 120 & -0.224 & -0.502 & 0.851 & -0.221 & -0.496 & 0.840 \\ 
        & 130 & 0.010 & -0.222 & 0.985 & 0.010 & -0.219 & 0.976 \\ 
        & 140 & 0.010 & 0.116 & 0.994 & 0.010 & 0.116 & 0.993 \\
        \hline
    \end{tabular}
    \caption{Values of the Stokes parameters obtained from the experimental data (under the "Measured" column) and from the results of applying our convex optimization method (Eqs.~\eqref{objective}-~\eqref{constraint2}) (under the "Corrected" column) for elliptically polarized light where the DOP $\geq$ 1 when calculated using the experimental data.}
    \label{tab:cp_data}
\end{table}
The results of our program, shown in Code 1 \cite{leamer_git}, applied to four sets of measured coherency matrices of linearly polarized light and four sets of measured coherency matrices of elliptically polarized light are respectively displayed in Fig.~\ref{fig:lp_fig} and Fig.~\ref{fig:cp_fig}.  Additionally, we present the values of the Stokes parameters for the points where the DOP $\geq$ 1 when calculated from the experimental data for linearly polarized (Table~\ref{tab:lp_data}) and elliptically polarized light (Table~\ref{tab:cp_data}).  The "Measured" column contains the measured values, while the "Corrected" column contains the values obtained by solving our minimization problem ~\eqref{objective}--~\eqref{constraint2}. In the non-physical cases where the DOP of $\mathbf{J}_\mathrm{measured}$ is greater than 1, the DOP of  $\mathbf{J}_{\mathrm{corrected}}$ obtained from our method is exactly 1. In the cases where $\mathbf{J}_\mathrm{measured}$ is physical, the obtained $\mathbf{J}_{\mathrm{corrected}}$ is equal to $\mathbf{J}_\mathrm{measured}$.  In both measurement schemes shown in Fig.~\ref{fig:setup_polarimetry}, we found that measuring the vertically polarized light through a QWP gave more non-physical results than that of a HWP. \par
We also found  $\mathbf{J}_{\min, \max}$ with the minimum and maximum DOP given a tolerance parameter of $\epsilon = 0.1$, which was done by solving the optimization problems defined in Eqs.~\eqref{min_dop} and ~\eqref{max_dop}, respectively. In most cases, the maximum DOP is found to be 1 and the minimum DOP is a constant value, which depends on $\epsilon$, lower than the measured DOP. Given the constraints in Eqs.~\eqref{min_dop} and ~\eqref{max_dop}, this is to be expected. There are a few exceptional points. In Fig.~\ref{fig:cp_fig}(c) at $\theta = 30^{\circ}, 40^{\circ}$, the maximum DOP is lower than 1. This is caused by the constraint that $\mathbf{J}_{\max}$ be $\epsilon$-close to $\mathbf{J}_{\mathrm{measured}}$, which makes the Stoke vector for $\mathbf{J}_{\max}$ lie inside the Poincar\'{e} sphere. Another exceptional case is demonstrated by the missing points on both the max DOP and min DOP plots in Fig.~\ref{fig:cp_fig}(a) at $\theta = 70^{\circ}$. Here, due to the same constraint above, the vectors corresponding to both $\mathbf{J}_\mathrm{max}$ and $\mathbf{J}_\mathrm{min}$ lie outside the Poincar\'{e} sphere, and thus, no solutions for both Eqs.~\eqref{min_dop} and ~\eqref{max_dop} can be found. 

\begin{figure}
    \centering
    \includegraphics[width=1\linewidth]{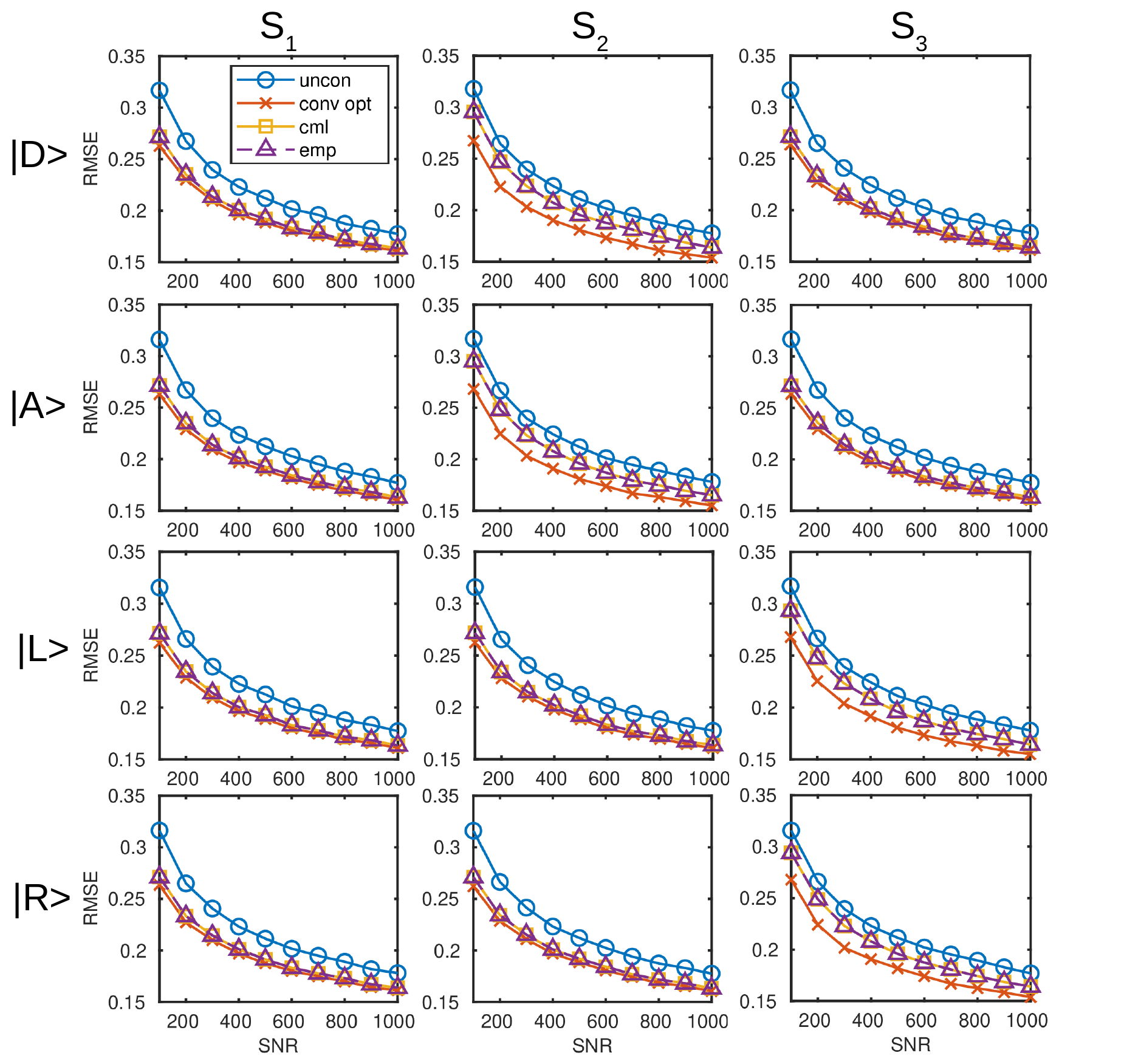}
    \caption{RMSE for the values of the Stokes parameters obtained from the unconstrained data (`uncon', blue circles), our convex optimization method (Eq.~\eqref{convex}, `conv opt', red crosses), the CML method (`cml', yellow squares), and the empirical method (`emp',purple triangles) as a function of the SNR \cite{hu_cml_2013}.  Each row of the figure corresponds to the results obtained by applying these methods to a set of data that was randomly generated using a different mean polarization state.  The results for the $s_1$, $s_2$, and $s_3$ parameters are grouped by column for each data set.}
    \label{fig:rmse}
\end{figure}

The normalized Stokes vectors corresponding to $\mathbf{J}_{\mathrm{corrected}}$ and $\mathbf{J}_{\mathrm{measured}}$ for each of the data points are displayed in the right column of Fig.~\ref{fig:lp_fig} for linearly polarized light and in the right column of Fig.~\ref{fig:cp_fig} for elliptically polarized light.  In every case, the vectors for $\mathbf{J}_{\mathrm{corrected}}$ and $\mathbf{J}_{\mathrm{measured}}$ are parallel. In the cases where the vector of $\mathbf{J}_{\mathrm{measured}}$ is outside the Poincar\'{e} sphere, the vector of $\mathbf{J}_{\mathrm{corrected}}$ ends on the surface of the Poincar\'{e} sphere.  This indicates that our method is successful at preserving the direction of the measured Stoke vectors while correcting for experimental errors. \par
 Another method for correcting the non-physical results in polarimetry experiments that is similar to the convex optimization method we have presented thus far is the constrained maximum likelihood (CML) estimator \cite{hu_cml_2013}.  In this work, Hu and Goudai develop an optimization scheme based on the Lagrange multiplier technique and apply it to correcting non-physical results obtained from the measurement of four points on the Poincar\'{e} sphere.  Additionally, the ``empirical" estimator method is presented in \cite{hu_cml_2013}. This method is applied by dividing the measured $s_1$, $s_2$, and $s_3$ parameters by the unconstrained DOP to obtain a set of new Stokes parameters that have a DOP of 1. \par

In Fig.~\ref{fig:rmse}, we compare the CML and empirical methods with our convex optimization technique by considering the estimation accuracy as a function of the signal-to-noise ratio (SNR) for the $s_1$, $s_2$, and $s_3$ parameters. Let $\ket{D}$, $\ket{A}$, $\ket{L}$, and $\ket{R}$ denote the Stoke vectors for diagonal, antidiagonal, left, and right polarizations, respectively (see also the labels in the Poincar\'e spheres in Figs.~\ref{fig:lp_fig} and ~\ref{fig:cp_fig}). For every value of SNR and for every polarization state $\ket{pol} \in \left\{ \ket{D}, \ket{A}, \ket{L}, \ket{R} \right\}$, we generate 50,000 samples of the Stokes vector $\ket{pol}$ contaminated by the additive Gaussian noise with zero mean and variance $\sigma^2 = \frac{1}{\mathrm{SNR}}$. The SNR is chosen to be in the range of 100 to 1000, where most polarimeters operate. The  accuracy of a polarimetry method is characterized via the root mean square error (RMSE). The RMSE between the reconstructed values for the Stokes parameter $s_k^\mathrm{est}$ and the known true value $s_k^0$ is given by
\begin{equation}
    \mathrm{RMSE}(s_k^{\mathrm{est}}) = \sqrt{\left\langle s_k^{\mathrm{est}} - s_k^0 \right\rangle ^2}, \qquad k=1,2,3,
\end{equation}
where $\langle \cdot \rangle$ represents the average over 50,000 samples. RMSE as a function of the SNR for each method and the unconstrained data are shown in Fig.~\ref{fig:rmse}. In every case, our convex optimization scheme outperforms the CML and empirical estimator methods albeit with a minor improvement in RMSE ($\sim 0.05$ at most). This would seem to imply that these methods are largely interchangeable, but our convex optimization scheme does have the advantage of being easier to implement as well as supporting a variety of noise models by utilizing different matrix norms.

\section{Conclusion}

We presented the convex optimization methods for the purpose of robust polarimetry as described in Sec.~\ref{sec:method}. We have demonstrated the validity of these methods using the experimentally measured results obtained for different polarization states and via different polarimetry schemes described in Sec.~\ref{sec:setup}. The performance of the developed techniques are discussed in Sec.~\ref{sec:results}. The presented methods do not depend on any \textit{a priori} information or calibration of the components nor on the type of experimental noise or error, and can be easily integrated into the post-processing of many polarimetry protocols.

\section*{Funding}

Defense Advanced Research Projects Agency (DARPA)(D19AP00043); U.S. Office of Naval Research (N000141912374); Louisiana Board of Regents’ Graduate Fellowship Program.

\section*{Acknowledgments}

This work was supported by the Defense Advanced Research Projects Agency (DARPA) grant number D19AP00043 under mentorship of Dr. Michael Fiddy. The views and conclusions contained in this document are those of the authors and should not be interpreted as representing the official policies, either expressed or implied, of DARPA, or the U.S. Government. The U.S. Government is authorized to reproduce and distribute reprints for Government purposes notwithstanding any copyright notation herein. RTG also acknowledges funding from the U.S. Office of Naval Research under grant number N000141912374. J.M.L. was supported by the Louisiana Board of Regents’ Graduate Fellowship Program.

\section*{Disclosures}

The authors declare no conflicts of interest.

\newpage

\bibliography{literature}

\end{document}